\documentclass{pasj00}

\begin{document}
\SetRunningHead{Y. Takeda}{Spectroscopic detection of stellar differential rotation}
\Received{2019/05/28}
\Accepted{2019/11/01}

\title{
On the detection of stellar differential rotation\\ 
based on the Fourier transform  of spectral line profiles
}

\author{
Yoichi \textsc{Takeda}\altaffilmark{1,2}
}
\altaffiltext{1}{National Astronomical Observatory, 2-21-1 Osawa, Mitaka, Tokyo 181-8588, Japan}
\email{takeda.yoichi@nao.ac.jp}
\altaffiltext{2}{SOKENDAI, The Graduate University for Advanced Studies, 
2-21-1 Osawa, Mitaka, Tokyo 181-8588, Japan}



\maketitle

\begin{abstract}
It is known that stellar differential rotation can be detected by 
analyzing the Fourier transform of spectral line profiles, since
the ratio of the first- and second-zero frequencies is a useful indicator. 
This approach essentially relies on the conventional formulation that 
the observed flux profile is expressible as a convolution of the rotational 
broadening function and the intrinsic profile, which implicitly assumes that 
the local intensity profile does not change over the disk. Although this 
postulation is unrealistic in the strict sense, how the result is affected 
by this approximation is still unclear. 
With an aim to examine this problem, flux profiles of several test lines 
(showing different center-to-limb variations) were simulated using a model 
atmosphere corresponding to a mid-F dwarf by integrating the intensity profiles 
for various combinations of $v_{\rm e}\sin i$ (projected rotational velocity), 
$\alpha$ (differential degree),
and $i$ (inclination angle), and their Fourier transforms were computed to check 
whether the zeros are detected at the predicted positions or not. 
For this comparison purpose, a large grid of standard rotational broadening 
functions and their transforms/zeros were also calculated.
It turned out that the situation critically depends on $v_{\rm e}\sin i$:
In case of $v_{\rm e}\sin i \gtsim 20$~km~s$^{-1}$ where rotational broadening 
is predominant over other line broadening velocities (typically several km~s$^{-1}$),
the first/second zeros of the transform are confirmed almost at the expected positions.
In contrast, deviations begin to appear as $v_{\rm e}\sin i$ is lowered,  
and the zero features of the transform are totally different from
the expectation at $v_{\rm e}\sin i$ as low as $\sim 10$~km~s$^{-1}$, 
which means that the classical formulation is no more valid.   
Accordingly, while the zero-frequency approach is safely applicable to studying 
differential rotation in the former broader-line case, it would be difficult 
to practice for the latter sharp-line case. 
\end{abstract}

%


\section{Introduction}

It has been known since olden days (Huang 1961) that stellar differential 
rotation may be detected by analyzing the profiles of spectral lines.
However, given the necessity of distinguishing very subtle difference of 
line shapes, early trials failed (Gray 1977, 1982) even by applying the Fourier 
transform technique, which is effective for extracting delicate information 
from line profiles (such as discrimination of rotation and turbulence; see, 
e.g., Gray 1988, 2005), It was at long last in the beginning this century 
when Reiners, Schmitt, and K\"{u}rster (2001) finally reported the existence 
of Sun-like differential rotation in $\psi$ Cap (F5 dwarf) based on the canonical
approach of comparing/fitting the transforms of observed and theoretical profiles 
for two blend-free lines (Fe~{\sc i}~5775 and Si~{\sc i}~5772) in the Fourier domain. 
Successively, Reiners and Schmitt (2002) found that the ratio of the frequencies at the 
first and second zeros in the Fourier transform of the broadening function is a good 
indicator of differential rotation, which provided a prospect of much easier model-independent 
detection by just measuring this ratio in the observed transform. Reiners and Schmitt 
(2003) further developed a method of extracting the rotational broadening function 
($G$) from a spectrum of several tens \AA\ portion by applying the Least Squares 
Deconvolution technique with the help of an appropriately adjusted template spectrum, 
by which the required first and second zeros can be effectively derived by 
transforming the resulting $G$ of sufficiently high-S/N ratio. Since then, 
making use of these advantages, Reiners and his collaborators extensively examined 
the nature of differential rotation in A-, F, and G-type stars of slow as well as 
rapid rotators (see, e.g.,  Ammler-von Eiff \& Reiners 2012, and the references therein).

Although their contributions in this field are regarded as significant, some concern
still remains regarding the assumption upon which their analysis is founded. 
That is, their adopted methodology (use of the frequency ratios of first and second zeros
in the Fourier domain, which are measured from the transform of the broadening 
function extracted by applying the deconvolution technique) is essentially based on
the postulation that the observed flux spectrum $F(\lambda)$\footnote{
For the case of line profile analysis, Fourier transform is usually applied 
to $1 - F(\lambda)/F^{\rm cont}$ (line depth relative to the local continuum) 
rather than $F(\lambda)$ itself, Accordingly, it should be kept in mind that 
this meaning is implicitly contained in $F(\lambda)$ used here. 
} is expressed
by a convolution of the stellar unbroadened spectrum $F_{0}(\lambda)$ and
the rotational broadening function $G(\lambda)$:
\begin{equation}
F(\lambda) = F_{0}(\lambda) \otimes G(\lambda),
\end{equation}
where $\otimes$ means ``convolution.'' 
By Fourier transforming these quantities, such as 
\begin{equation}
f(\sigma) \equiv \int_{-\infty}^{+\infty}F(\lambda) \exp(2\pi i \lambda \sigma) {\rm d}\lambda,
\end{equation}
we have
\begin{equation}
f(\sigma) = f_{0}(\sigma) g(\sigma),
\end{equation}
which describes the advantage of Fourier technique; i.e., since ``convolution'' turns into
simple ``product'' in the Fourier domain, the zeros of $g(\sigma)$ are simply inherited
to $f(\sigma)$ to be directly observable.  

It should be noted, however, that equation~(1) is nothing but an assumption.
Since observed flux is an integration of specific intensities over the stellar disk 
(whose spectra generally differ from position to position), the necessary condition 
for equation (1) to strictly hold is that $I_{0}(\lambda)/I_{0}^{\rm cont}$ (line 
shape of the residual intensity normalized by the continuum) does not vary over the disk; 
in this case, the residual flux $F_{0}(\lambda)/F_{0}^{\rm cont}$ can be equated to 
the position-independent $I_{0}(\lambda)/I_{0}^{\rm cont}$ (see, e.g., Gray 2005).
Unfortunately, since this is an unrealistic condition (even the classical line-formation 
theory such as the Milne--Eddington model predicts a result against it; e.g.. 
Eq.~(10-29) in Mihalas 1978), characteristic center-to-limb variation in the 
strength/profile of any line must more or less exist. 

Generally, in case of $v_{\rm e}\sin i$ (projected rotational velocity) determination, 
which is reflected by the line width (1st order effect), significant errors would 
not be expected even if equation~(1) does not hold. However, detection of 
differential rotation ($\alpha$) is a far more difficult and delicate task, which 
requires discriminating very subtle difference in the profile shape (2nd order effect).
Thus, whether this assumption is valid or not may bring about crucial impact on 
the results of analysis, which should be clarified by all means. 

Nevertheless, it has been barely investigated so far, to our knowledge, how much 
deviation from the profile constancy would be expected case by case or how much 
error would result by using equation~(1) in the line-profile analysis. That is, 
investigators in this field have invoked this assumption without any convincing
verification. Perhaps the main reason may be the insufficiency of information  
regarding how the profiles of different lines exhibit changes over the stellar disk.

In order to improve this situation, Takeda and UeNo (2017a, 2017b, 2019) recently conducted
a comprehensive spectroscopic study to clarify the behaviors of the strengths and 
broadening widths across the solar disk for a number of spectral lines, and arrived 
at a reasonable understanding of their center-to-limb variations in terms of the
line properties (e.g., species, excitation potential, strength, etc.). It is, therefore,
now possible to guess the tendency of center--limb variations of line profiles
at least for solar-type stars. 

Accordingly, the purpose of this study is to examine the validity of the method for 
detecting stellar differential rotation, which makes use of the zeros of Fourier 
transforms and is founded on the assumption of equation~(1). 
This verification test consists of the following procedures:
\begin{itemize}
\item
Model (flux) profiles of several representative test lines are simulated for various 
combinations of rotation parameters by integrating 
the intensity profiles at various points of the stellar disk, while taking into account 
the center--limb variations of local line profiles as well as continuum 
intensities (i.e., limb darkening).     
\item
For each of the resulting flux profiles, Fourier transforms are computed, 
in which the positions of zero frequencies are then searched for.  
If the conventional assumption really holds, the first and second zeros ($\sigma_{1}$ 
and $\sigma_{2}$) should appear at the same positions as those given by $g(\sigma)$
according to equation~(3); in this case, the approach adopted by Reiners et al. 
is justified. Otherwise, the differential rotation judged by the $\sigma_{2}/\sigma_{1}$ 
ratio can not be regarded as trustworthy any more. 
\end{itemize}

In this way, this test would serve as a touchstone for the validity or applicability 
limit of the detection method of differential rotation by using $\sigma_{2}/\sigma_{1}$.
As a prerequisite, however, it is necessary to know the precise positions 
of zeros in $g(\sigma)$ (Fourier transform of the rotational broadening function 
$G(\lambda)$) to be compared with the simulated ones, 
which is addressed first in the next section.   

The remainder of this paper is organized as follows. Following section~2 (for the 
rotational broadening function), details of line-profile modeling are spelled out in 
section~3, where the selected 8 test lines, adopted angle-dependent micro/macro-turbulence, 
and simulations of line profiles as well as their Fourier transforms are described. 
Discussions of the results (limb-darkenig coefficients, behaviors of zero frequencies, 
interpretation of previous studies, and how to deal with the slow-rotator case) are 
presented in section~4, which is followed by section~5 (summary and conclusion). 
In addition, two supplemantary appendices are presented: Appendix~1 describes the 
behaviors of solar macro- and micro-turbulence based on the results of Takeda and 
UeNo (2019), which serve as the basis of the functional forms of the turbulences 
adopted in this study. Appendix~2 shows the results of test calculations, which 
demonstrate the significant impact of angle-dependence in turbulent velocities 
on the main conclusion.    

\section{Rotational broadening function}

Regarding the rotational broadening function $G$ to be used in equation~(1), 
it can be evaluated only numerically when a star rotates differentially 
(unlike the case of rigid rotation where $G$ is analytically expressible).
Although this was tried by several earlier studies in 1980s (Bruning 1981, 1982;
Garc\'{\i}a-Alegre, V\'{a}zques, \& W\"{o}hl 1982), Reiners and Schmitt (2002) 
computed this function more extensively for various combinations of the parameters.
However, since these previous investigations employed a simple method of integrating 
the profile of a test line over the stellar disk (by Doppler shifting according 
to the local line-of-sight velocity), such computed $G$ is not exact but more 
or less approximate, because the test line can not be a $\delta$-function but
has a finite width. In this study is adopted an alternative method of using iso-velocity
contours (cf. Huang 1961), by which precise evaluation of $G$ is possible.
  
Let two coordinate systems having the common origin and common $x,x'$-axis be defined, 
$(x, y, z)$ and $(x',y',z')$, where the $y$-axis of the former is the line of sight
and the $z'$-axis of the latter is the rotation axis of a star, both making
an inclination angle $i$ (cf. figure~1).
As usually done, the rotation law of a (spherical) star is assumed to be 
\begin{equation}
\omega (l) = \omega_{\rm e} (1 - \alpha \sin^{2} l),  
\end{equation}
where $\omega$ is the angular velocity at a latitude $l$, 
$\omega_{\rm e}$ is the equatorial angular velocity, and $\alpha$ is the parameter
indicating the degree of differential rotation.\footnote{
This form is widely used in most stellar applications, which was devised
in analogy with the solar differential rotation. In the solar case, however,
the $\sin^{4}l$ term is often included in addition to the $\sin^{2}l$ term
(see, e.g., Sect.~4,2 of Takeda \& Ueno 2011)
}
Then, in the $(x',y',z')$ system, the line-of-sight velocity of a stellar surface 
element as viewed along the $y'$ direction is written as
\begin{equation}
v_{y'} = x' \omega_{\rm e} (1 - \alpha z'^{2}),
\end{equation}
where the position coordinates are expressed in unit of the stellar radius.
Further, measuring the velocity in unit of the equatorial rotation velocity 
by introducing the non-dimensional parameter $k$ $(\equiv v/v_{\rm e})$
this equation reduces to
\begin{equation}
k' = x' (1 - \alpha z'^{2}). 
\end{equation}
In the $(x,y,z)$ system, this relation is transformed as
\begin{equation}
k = x [1 - \alpha (-\sqrt{1-x^{2}-z^{2}} \cos i + z \sin i)^{2}],
\end{equation}
which defines the iso-velocity trajectory $c_{k}(x,z)$ on the $x$--$z$ plane
corresponding to any given velocity $k$ ($-1 \le k \le 1$).\footnote{
For the purpose of determining iso-velocity contours, it is not necessarily 
practical to directly use this equation~(7), because $x$ can not be described 
by an analytical function of $z$, $\alpha$, and $i$ (though $z$ is expressed 
analytically in terms of $x$, $\alpha$, and $i$).
It is much easier to start from equation~(6), which defines a closed ``loop'' 
of iso-velocity on the stellar sphere, in the sense that all surface points on 
this loop  (symmetric with respect to the $x'$--$z'$ plane) have the same 
line-of-sight velocity as seen along the $y'$-direction. Here, the bottom line is 
that this 3-dimensional loop on the sphere retains its iso-velocity character also in 
the $(x ,y ,z)$ system (i.e., as seen along the $y$-direction). Then, what should be 
done is just to transform the $(x',y',z')$ coordinates of all the points on this loop 
into the $(x ,y ,z)$ system. The resulting ($x$, $z$) coordinates simply provide for 
the iso-velocity contour, though it has to be kept in mind that the points with $y > 0$  
should not be included (because they are on the invisible side of the sphere).   
}
Some examples of $c_{k}(x,z)$ contours for selected representative cases 
($\alpha = -0.4, 0.0, +0.4$ and $i$ = 10, 50, 90$^{\circ}$) are depicted in figure~2.   
Then, $G(k)$ (rotational broadening function at $k$) is calculated by
integrating the visible area of iso-velocity band (whose width is
proportional to $(\partial k/\partial x)^{-1}$; i.e., breadth in $x$ per unit $k$) 
while taking into account the limb-darkening effect (assumed to follow the linear law)
\begin{equation}
I(\theta) = I_{0}(1 - \epsilon + \epsilon \cos\theta),
\end{equation}
where $\epsilon$ is the limb-darkening coefficient, $I_{0}$ is the intensity at the disk 
center, and $\theta$ is the angle of the outgoing ray relative to the surface normal. That is,  
\begin{equation}
G(k) \propto \int_{c_{k}(x,z)}^{}  (\partial k/\partial x)^{-1} 
    (1 - \epsilon + \epsilon \sqrt{1 - x^{2} - z^{2}}) {\rm d}z,
\end{equation}
where the integration is carried out along the contour $c_{k}(x,z)$ corresponding to $k$.

The computations of $G(k)$ (for 201 points from $k = 0$ to 1 with an increment of 0.005)
were done for 4158 cases with combinations of 21 values of $\alpha$ ($-5.0$, $-4.5$, $-4.0$,
$\cdots$, +4.5, +5.0), 18 values of $i$ (5$^{\circ}$, 10$^{\circ}$, 15$^{\circ}$, $\cdots$, 
85$^{\circ}$, 90$^{\circ}$), and 11 values of $\epsilon$ (0.0, 0.1, 0.2, $\cdots$, 0.9, 1.0).
The resulting data of these broadening functions $G(k)$ are presented as online materials
(each of the ``gprofs\_a????e???.dat'' files). Then, their Fourier transforms $g(q)$\footnote{
In this study, non-dimensional Fourier frequency (corresponding to the non-dimensional
variable $k$ in the real space) is denoted as $q$, distinguished from 
the frequency $\sigma$ used in equation~(2) which has the dimension of wavelength$^{-1}$. 
} were then computed, and the zero frequencies as well as the heights of sidelobes
were further measured, as summarized in the online data table ``g\_transform\_info.dat''.

It may be worth comparing the resulting values of the key quantities known as indicators 
of differential rotation, such as $q_{2}/q_{1}$ (second-to-first zero frequency ratio)
or $I_{1}/I_{2}$ (first-to-second sidelobe height ratio), with those derived by Reiners
and Schmitt (2002). Such a comparison is illustrated in figures~3a--3d, which are 
confirmed to be reasonably consistent with Fig.~5 and Fig.~7 of their paper.
Likewise, figures~3e and 3f show that $\alpha/\sqrt{\sin i}$ can be approximated
by analytical functions of $q_{2}/q_{1}$, as argued by Reiners and Schmitt (2003)
[cf. their Fig.~11 along with their equations (5) and (6)].    

\section{Modeling of spectral line profiles} 

\subsection{Assumed stellar parameters}

Since Reiners and his colleagues have investigated the nature of differential rotation 
mainly for stars around F-type (i.e., late-A through early-G dwarfs), this study 
focuses on a F5 dwarf as a representative example, for which the following stellar parameters
may be assigned by consulting Table~B.1 of Gray (2005). $T_{\rm eff} = 6500~K$ (effective
temperature), $M = 1.4~M_{\odot}$ (stellar mass), $R = 1.4~R_{\odot}$ (stellar radius),
$\log g = 4.29$ (surface gravity in cgs unit, derived from $M$ and $R$), and 
[Fe/H] = 0.0 (solar metallicity). The model atmosphere used for this study was generated 
by interpolating Kurucz's (1993) ATLAS9 grid of solar metallicity models.

\subsection{Center-to-limb variations of solar spectral lines}

Since the line profiles of an F5 star are to be simulated in the solar analogy,
the center--limb variations in the line strengths/widths over the solar disk should make
an important basis, for which the consequences of Takeda and UeNo (2017a, 2017b, 2019) are 
briefly summarized below.
\begin{itemize}
\item
Regarding the macroturbulence, the frequently used radial-tangential model was
concluded to be inadequate, while the classical Gaussian model being more preferable.
The solar Gaussian macroturbulence ($v_{\rm mac}$) moderately increases toward the limb from 
$\sim$~1--2~km~s$^{-1}$ (center) to $\sim$~2--3~km~s$^{-1}$ (limb), though some dependence 
upon the line strength is also observed (cf. Fig.~9 in Takeda \& UeNo 2017a, where
this macroturbulence is denoted as $V_{\rm los}$). See also figure~9 in appendix~1. 
\item
The trends of center-to-limb variations in the equivalent widths are diversified 
case by case (mainly depending upon the temperature sensitivity), which are classified 
in terms of (a) whether the species is of minor population or major population 
(determined by the ionization potential), (b) excitation potential, and (c) line strength 
(i.e., saturation degree) (Takeda \& UeNo 2017b, 2019).
\item
The logarithmic change of the equivalent width ($W$) relative to the disk center 
value ($W_{0}$) is in the range of $|\log (W/W_{0})| \ltsim 0.3$, whichever 
a line is strengthened or weakened toward the limb (Takeda \& UeNo 2017b, 2019).
\item
An angle-dependent microturbulence ($v_{\rm mic}$) (i.e., increasing toward the limb)
has to be introduced (except for unsaturated very weak lines), in order to reproduce 
such established trend of center--limb variation in $W$, which can be concluded by
analyzing Takeda and UeNo's (2019) $W$ data (such as done by Holweger, Gehlsen, \& Ruland 
1978). See also figure~10 in appendix~1.
\end{itemize}

\subsection{Choice of turbulence parameters}

By consulting the case for the Sun summarized in the previous subsection,  
the $v_{\rm mac}$ and $v_{\rm mic}$ to be used for an F5 star were assigned as follows, 
while postulating that (i) solar values are scaled by taking into account the differences of 
atmospheric parameters and (ii) similar angle-dependence (i,e., functional form of $\theta$) holds.

The empirical formula for $v_{\rm mic}$ of F--G stars derived by Takeda et al. 
(2013; cf. equations (1) and (2) therein) yields  
$v_{\rm mic} = 1.07$~km~s$^{-1}$ (Sun; $T_{\rm eff}$ = 5780~K, $\log g$ = 4.44) and
$v_{\rm mic} = 1.76$~km~s$^{-1}$ (F5 dwarf; $T_{\rm eff}$ = 6500~K, $\log g$ = 4.29).
Accordingly, the scaling relation of $v^{\rm F5}/v^{\odot} \simeq 1.6$ is assumed for both 
$v_{\rm mac}$ and $v_{\rm mic}$.
Although this scaling factor was derived from the $v_{\rm mic}$ vs. 
$T_{\rm eff}$ relation, it is almost consistent with the similar 
trend for $v_{\rm mac}$. Regarding the graphical overview of the published 
$T_{\rm eff}$-dependent relations, see Fig.~7 (left panel) of Ryabchikova 
et al. (2016) and Fig.~1d of Takeda, Hashimoto, and Honda (2017) for 
$v_{\rm mic}$, while Fig.~7 (right panel) of Ryabchikova et al. (2016) 
and Fig.~14 of Takeda and UeNo (2017a) for $v_{\rm mac}$.

By examining the results of $v_{\rm mac}$ (Gaussian macroturbulence) derived by Takeda 
and UeNo (2017a) for the Sun, $v_{\rm mac}^{\odot}$ may be roughly approximated by the relation\footnote{
It turned out that the use of $\sin\theta$ is better (than using $\cos\theta$), if the $\theta$-dependence 
is to be represented by such a simple linear-term relation.
} 
\begin{equation}
v_{\rm mac}^{\odot}(\theta) = 1.5 + 1.0 \sin\theta \;\;\; ({\rm km}\;{\rm s}^{-1}),
\end{equation} 
which further yields 
\begin{equation}
v_{\rm mac}^{\rm F5}(\theta) = 2.4 + 1.6 \sin\theta \;\;\; ({\rm km}\;{\rm s}^{-1})
\end{equation}  
by applying the scaling relation.

Regarding the solar microturbulence $v_{\rm mic}^{\odot}$, Holweger, Gehlsen, and Ruland 
(1978) derived 1.0~km~s$^{-1}$ (disk center) and 1.6~km~s$^{-1}$ (near to the limb 
at $\cos\theta = 0.3$ or $\sin\theta = 0.95$). Then, by assuming the angle-dependence similar to 
that of macroturbulence, we may write
\begin{equation}
v_{\rm mic}^{\odot}(\theta) = 1.0 + 0.6 \sin\theta \;\;\; ({\rm km}\;{\rm s}^{-1}),
\end{equation} 
which is further scaled as  
\begin{equation}
v_{\rm mic}^{\rm F5}(\theta) = 1.6 + 1.0 \sin\theta \;\;\; ({\rm km}\;{\rm s}^{-1}).
\end{equation}  

The center-to-limb behaviors of the finally adopted macro- and micro-turbulences for the F5 star, 
$v_{\rm mac}^{\rm F5}(\theta)$ and $v_{\rm mic}^{\rm F5}(\theta)$ expressed by equations (11) 
and (13), are graphically shown in figure~4a.

\subsection{Adopted test lines}

As mentioned in subsection~3.2, the center-to-limb variations of equivalent widths 
differ from line to line depending on the line properties. After some trial calculations,
8 fictitious lines (Fe~{\sc i} lines of $\chi_{\rm low}$ = 0, 3, and 6~eV; O~{\sc i} line
of $\chi_{\rm low}$ = 9~eV; two kinds of line strengths for each case with $W_{0}$ = 30~m\AA\ 
and 100~m\AA; line center wavelength at 5000~\AA) were decided to adopt (cf. table~1), 
which were so selected as to reproduce such a diversified center--limb variations
in the range of $|\log (W/W_{0})| \ltsim 0.3$ as shown by the Sun.
The resulting $W(\theta)$ values for each line, which were computed by Kurucz's (1993) 
WIDTH9 program with the angle-dependent microturbulence $v_{\rm mic}^{\rm F5}(\theta)$ 
[cf. equation~(13)], are plotted against $\cos\theta$ in figure~4b,
In addition, the $W$ vs. $\cos\theta$ trends corresponding to a constant (i.e., 
angle-independent) $v_{\rm mic}$ of 1.6~km~s$^{-1}$ are also depicted in figure~4c
for comparison (which are related to the test calculations described in appendix~2).  

\subsection{Line profile calculation}

For the purpose of simulating line profiles of a differentially rotating star,
the modified version of the program code CALSPEC (Takeda, Kawanomoto, \& Ohishi 2008)
was used, which generates the stellar flux spectrum by integrating the specific 
intensities along the line of sight at each of the many small segments over the visible 
disk (computed with the corresponding $v_{\rm mic}$ from the model atmosphere 
and then broadened according to the relevant $v_{\rm mac}$).
Since only comparatively slow rotators are involved in this study, gravity darkening
as well as gravity distortion were neglected, which means that the star's shape
remains spherical, the continuum intensity (dependent only on $\theta$) is 
always circularly symmetric on the disk, and only one atmospheric model of
($T_{\rm eff}$, $\log g$) = (6500, 4.29) is relevant. 

Regarding the three parameters ($v_{\rm e}\sin i$, $\alpha$, and $i$) to be specified 
for simulating a line profile, calculations were done for combinations of 5 
values of $v_{\rm e}\sin i$ (10, 20, 30, 40, and 50~km~s$^{-1}$), 11 values of $\alpha$  
($-0.5$, $-0.4$, $-0.3$, $\cdots$, +0.4, +0.5), and 9 values of $i$ (10$^{\circ}$, 20$^{\circ}$, 
30$^{\circ}$, $\cdots$, 80$^{\circ}$, 90$^{\circ}$). The profiles of 8 lines (table~1) were 
computed for all these cases in the wavelength range of [4998~\AA, 5002~\AA] with an increment 
of 0.005~\AA\ (i.e., 801 points within $\pm 2$~\AA\ from the line center).  

Then, the Fourier transforms were computed for the line depths [$1- F(\lambda)/F^{\rm cont}$] 
of the resulting profiles, and the positions of zero frequencies ($\sigma_{1}$, $\sigma_{2}$, 
$\cdots$) were measured. Since these $\sigma$'s have dimensions of wavelength$^{-1}$,
they were transformed into non-dimensional frequencies ($q_{1}$, $q_{2}$, $\cdots$)
by the relation 
\begin{equation}
q_{j} = \sigma_{j} \lambda (v_{\rm e}\sin i)/c \;\;\; (j = 1, 2, \cdots)
\end{equation}
($\lambda$: wavelength, $c$: speed of light),
which are independent of $v_{\rm e}\sin i$ and directly comparable with
$q$'s of the broadening function $G(k)$ derived in section~2.

\section{Discussion}

\subsection{Limb-darkening coefficient}

As a preparation for comparing the Fourier transform properties (i.e., zero frequencies) 
of simulated line flux profiles (computed by integrating the local model-based intensity 
profiles over the disk) with those of the rotational broadening function 
[$G(k; i, \alpha, \epsilon)$], it is necessary to adequately specify the value of 
$\epsilon$ (limb-darkening coefficient) compatible with the calculated model profiles. 
Although in the papers of Reiners group (e.g., Ammler-von Eiff \& Reiners 2012) was varied 
this parameter between 0 and 1 (around the standard value of 0.6) 
to estimate the uncertainty range of $q_{2}/q_{1}$ as if it was unknown, 
its value relevant for the atmospheric parameters 
(especially $T_{\rm eff}$) and line wavelength adopted in this study was evaluated 
from the model atmosphere. 
Figures~5a--5c show the spectral distribution (of the specific intensity $I_{\nu}$ 
at the disk center computed by ATLAS9 program), $I_{\nu}$ vs. $\cos\theta$ relations 
(along with the linear-regression lines) at representative wavelengths, and the 
resulting $\epsilon$ values plotted against wavelength, respectively. In the present 
case where the wavelengths of fictitious lines were assumed to be 5000~\AA, 
figure~5c indicates that $\epsilon = 0.6$ is a reasonable choice.

\subsection{Behaviors of zero positions}

Figure~6 demonstrates how the zero frequencies ($q_{1}$, $q_{2}$, and $q_{3}$) measured 
from the transforms of the simulated profiles (cf. subsection~3.5), which were computed 
for the 2600c30w100 case with 3 values of $v_{\rm e}\sin i$ (10, 20, and 30~km~s$^{-1}$),\footnote{
Although the actual calculations were done for five values of $v_{\rm e}\sin i$  
(10, 20, 30, 40, and 50~km~s$^{-1}$), only the first three cases are discussed 
in this paper, because the results for 40 and 50~km~s$^{-1}$ turned out to 
show a similar tendency as that for 30~km~s$^{-1}$.} 
5 values of $\alpha$ ($-0.4$, $-0.2$, 0.0, +0.2, and +0.4), and 9 values of $i$ 
(10$^{\circ}$, 20$^{\circ}$, $\cdots$, and 90$^{\circ}$), are compared with 
those corresponding to the standard rotational broadening function for 
$\epsilon = 0.6$ (cf. section~2). Although this figure is 
for the 2600c03w100 case, the situations for the other 7 lines (cf. table~1) are 
almost similar and not much different.

It is apparent from figure~6 that whether the consistency is realized or not 
critically depends upon $v_{\rm e}\sin i$:\\
--- For the case of $v_{\rm e}\sin i = 30$~km~s$^{-1}$ (right panels),
a satisfactory agreement is observed for any of $q_{1}$, $q_{2}$, and $q_{3}$,
except for the low $i$ ($<50^{\circ}$) cases at $\alpha = +0.4$, where the 
transform $g(q)$ has problems in terms of zero detection (cf. the caption to figures~3).\\
--- However, regarding the $v_{\rm e}\sin i = 20$~km~s$^{-1}$ case (center panels), an 
appreciable deviation begins to appear in $q_{3}$ (especially for larger $\alpha$ 
values) though $q_{1}$ and $q_{2}$ are still in agreement.\\  
--- Finally at the $v_{\rm e}\sin i = 10$~km~s$^{-1}$ case (left panels), any consistency 
vanishes in the sense that the predicted zeros are not reproduced at all
by the transforms of simulated transforms.\\ 

Regarding the $q_{2}/q_{1}$ ratio, which has been used as a useful indicator
of differential rotation (cf. section~1), the relative differences of
$(q_{2}/q_{1})_{\rm mes}$ (measured from the transform of simulated profiles) and  
$(q_{2}/q_{1})_{\rm exp}$ (expected from $g(q)$) are plotted against 
$(q_{2}/q_{1})_{\rm exp}$ in figure~7 for each of the 8 lines. This figure confirms that 
the tendency described above (i.e., the critical role of $v_{\rm e}\sin i$) similarly holds 
for all the studied lines irrespective of the differences in center-to-limb variations.
It is also worth noting that the discrepancy is more likely to be seen for larger 
(positive) $\alpha$ in the transition case. For example, in the center and right panels 
($v_{\rm e}\sin i$ = 20 and 30~km~s$^{-1}$) of figure~7, the results for $\alpha = +0.4$ 
(represented by filled squares at lower $(q_{2}/q_{1})_{\rm exp}$ of $\sim$~1.2--1.4) 
show more deviations compared to the cases of other (smaller) $\alpha$. 

Consequently, the following conclusion may be drawn regarding the applicability of 
Fourier transform method for spectroscopically detecting stellar differential rotation.
\begin{itemize}
\item
In the realistic case where line-profiles are not constant over the disk,
the validity of equation~(1) depends upon the extent of $v_{\rm e}\sin i$
in comparison to the other broadening velocities (thermal velocity, $v_{\rm mic}$, 
and $v_{\rm mac}$; root-mean-square is $\sim 5$~km~s$^{-1}$ in the present case).  
\item
If the former is predominant over the latter (i.e., $v_{\rm e}\sin i \gtsim 30$~km~s$^{-1}$),
equation~(1) is practically valid and the $q_{2}/q_{1}$ ratios expected from the
rotational broadening function are reproduced in the transforms of actual profiles.
\item
However, if the latter is comparable (or not negligible) to the former
(i.e., $v_{\rm e}\sin i \ltsim 20$~km~s$^{-1}$), even a small deviation from the 
constancy can violate the expected zeros in the transform and thus 
equation~(1) does not hold any more.
\item
In addition, around the critical $v_{\rm e}\sin i$ ($\sim$~20~km~s$^{-1}$), the risk 
for the breakdown of equation~(1) tends to be higher as $\alpha$ increases.   
\end{itemize}

In a different perspective, an intuitive and qualitative interpretation of these 
trends may be possible. That is, the classical method is applicable if the line profile 
shows a characteristic shape of rotational broadening function (i.e., rounded {\bf U} shape),
which yields conspicuous sidelobes and clear zeros in the Fourier space.
On the other hand, if the profile has lost such a rotational shape to have
a rather sharp appearance (i.e., {\bf V} shape such as like a Gaussian function), 
zeros as well as sidelobes are hardly detected in the Fourier space, which means
that the procedure invoking zero frequencies does not have any chance.
Accordingly, if thermal/turbulent velocities are no more negligible compared to 
$v_{\rm e}\sin i$, Fourier transform method using $q_{2}/q_{1}$ would become ineffective 
because line profiles tend to have an undesirable {\bf V} shape. Likewise, in 
the transition case of critical $v_{\rm e}\sin i$ ($\sim 20$~km~s$^{-1}$), the same 
goes if $\alpha$ is sufficiently large (which makes profiles sharper; 
cf. Fig.~3 in Reiners \& Schmitt 2002).

\subsection{Impact on the results of previous studies}

In a number of papers published by Reiners and his collaborators was investigated 
the nature of differential rotation for various A-, F-, and G-type stars of a wide range
of stellar rotation ($v_{\rm e}\sin i$ from a few km~s$^{-1}$ to $\sim 300$~km~s$^{-1}$; 
see, e.g., Ammler-von Eiff \& Reiners 2012) by using the $q_{2}/q_{1}$ ratio.
It may be appropriate here to review the consequences established by them based on 
what has been elucidated in this study.

According to the conclusion described in subsection~4.2, their results would 
have no problems as long as stars showing moderate or large rotational broadening 
($v_{\rm e} \sin i \gtsim$~20--30~km~s$^{-1}$) are concerned (which constitute
most of their sample stars), because the observed $q_{2}/q_{1}$ ratio can be regarded 
as equivalent to that of the standard rotational broadening function (reflecting 
the degree of differential rotation).
Therefore, their first detection of differential rotation ($\alpha \simeq 0.15$) 
for $\psi$ Cap based on the Fourier transforms of Fe~{\sc i} 5775 
or Si~{\sc i} 5772 line profiles (cf. Reiners, Schmitt, \& 
K\"{u}rster 2001; Reiners \& Scimitt 2002) should be all right, since 
this star is sufficiently broad-lined ($v_{\rm e}\sin i = 42$~km~s$^{-1}$).

However, the situation is different for slow rotators ($v_{\rm e} \sin i \ltsim$~10--20~km~s$^{-1}$),
where the $q_{2}/q_{1}$ ratio is no more usable because the zero frequencies in the 
Fourier space considerably deviate from the positions expected from the broadening function 
(or even zeros do not appear at all). This means that differential rotation can not be
studied by this technique any more for such sharp-lined stars.\footnote{
What matters here is Reiners et al.'s comparatively later results related to 
low $v_{\rm e}\sin i$ stars, which were based on the broadening function ($G$) 
derived by applying the LSD technique to spectra of wide wavelength coverage 
(i.e., not from individual line profiles) as mentioned in section~1.
In the author's opinion, the consequence of this study specifically derived 
for single-line profile analysis equally applies to the wide-range spectrum 
analysis with the LSD method for the following reasons: (i) Since the proposition 
that ``equation~(1) is no longer valid for such slow rotators'' was proved for 
all 8 representative lines chosen to cover the various center--limb variations 
of most lines, it should apply to any spectral lines (or their aggregates). 
(ii) This means that the breakdown of equation~(1) at low $v_{\rm e}\sin i$ 
is expected also for wide-range spectra comprising many lines in general. 
(iii) As such, any results obtained by LSD for slow rotators should not be reliable,
because equation~(1) is the fundamental postulation of LSD (i.e., deconvolution).
In this respect, their important conclusion, such as that the 
fraction of solar-like differential rotators ($\alpha > 0$) tends to progressively 
increase with a decrease in $v_{\rm e}\sin i$ (amounting to $\sim 60$\% at 
$v_{\rm e}\sin i \sim$~10--20~km~s$^{-1}$; cf. Fig.~8 in Ammler-von Eiff \& 
Reiners 2012), should be regarded with caution and worth reinvestigation.
However, an anonymous referee of this paper objected to this view, 
arguing that (a) information derived for only 8 fictituous lines does not suffice 
to discuss the results obtained by LSD method because many lines of diverse  
properties are involved and (b) extensive calculation of realistic theoretical 
synthetic spectrum including a large number of lines and its inverse analysis 
by the LSD technique (just like Reiners et al. did) is necessary. Since such 
large-scale simulation is outside the scope of this paper, the author would 
take modest attitude of not concluding anything about the validity of
Reiners et al.'s results obtained by LSD for the moment, the clarification of 
which by future studies is awaited.   
}

In this context, it may be worth remarking that Reiners et al. also mentioned 
the applicability limit of their method concerning $v_{\rm e} \sin i$, 
in the sense that a critical value of (minimum) $v_{\rm e} \sin i$ 
exists, below which their Fourier transform method using zero frequency 
ratio is no more effective. However, what is meant by the $v_{\rm e}\sin i$ 
limit in their papers simply concerns the spectral resolution\footnote{
It appears that the term ``spectral resolution'' (denoted by $R$) is used 
in their papers to represent $\lambda/\Delta x$ (e.g., Reiners \& Schmitt 2002),
where $\Delta x$ is the sampling step (in the same unit as the wavelength 
$\lambda$), by which the Nyquist frequency is defined as 
$\sigma_{\rm N} \equiv 1/(2\Delta x)$. 
Accordingly, the difference from the usual meaning of 
$R \equiv \lambda/\Delta\lambda_{\rm FWHM}$ 
should be kept in mind, where $\Delta\lambda_{\rm FWHM}$ is 
the full-width at half-maximum of the instrumental profile 
(the relation $\Delta\lambda_{\rm FWHM} \sim$~2--3 $\Delta x$ 
typically holds in the normal use of the spectrograph).} 
(sampling step) of line profile data, on which Fourier transform is calculated.
That is, since the relevant zero frequencies ($\sigma_{1}$ and $\sigma_{2}$; 
containing information of rotation) progressively shift to higher
frequency range as $v_{\rm e}\sin i$ decreases, $\sigma_{2}$ is
no more measurable once it exceeds the Nyquist frequency ($\sigma_{\rm N}$;
determined by the sampling step) because of the aliasing problem, 
by which the limiting value of minimum $v_{\rm e}\sin i$ is defined.
For example, Reiners and Schmitt (2003; see Sect.~3 therein) 
states that ``To detect differential rotation on stars with 
$v_{\rm e}\sin i < 20$~km~s$^{-1}$, spectra with a resolution of 
$R \ge 100000$ are needed'' (see also Fig.~9 in Reiners \& Schmitt 2002).
In contrast, the consequence of this study is essentially different, 
which implies ``intrinsic'' inapplicability of the $\sigma_{2}/\sigma_{1}$ 
method for slow rotators of $v_{\rm e}\sin i < 20$~km~s$^{-1}$ (regardless 
of the sampling step or spectrum resolving power).

\subsection{Prospect of line profile study for slow rotators}

Then, how could the information of rotational features for low $v_{\rm e}\sin i$ stars 
($\ltsim$~10~km~s$^{-1}$) be extracted from their line profiles? Since equation~(1) does 
not hold and the information of rotational broadening function is useless in this case,
a universally applicable simple approach (such as the use of only $q_{2}/q_{1}$) should
not be counted upon any more. Presumably, there is no royal road but to carefully compare 
the observed profile with the theoretical ones simulated by the disk integration method 
(such as done in this study).

As a demonstration, the line profiles ($F(\lambda)/F^{\rm cont}$) and 
their Fourier transform amplitudes ($|r(q)|$) of the 2600c30w100 line 
computed for the $v_{\rm e}\sin i$ = 10~km~s$^{-1}$ case are illustrated 
in figure~8. It can be seen from this figure (right-hand panels) that the appearance of 
$r(q)|$ considerably changes by varying $\alpha$ as well as $i$, which means that 
studying Fourier transforms is still useful for separate determinations of these parameters
(even though the predictions from the broadening function are useless here).
Besides, the line flux profile itself would also be helpful, though its variations are 
generally subtle. For example, the residual flux at $\lambda \simeq 5000.1$~\AA\ 
($F(\lambda)/F^{\rm cont} \sim 0.7$) is a ``fixed point'' (i.e., invariable for changing 
$i$ or $\alpha$), as seen from the left-hand panels of figure~8, which may be used to 
establish $v_{\rm e}\sin i$ without prior knowledge of $\alpha$ and $i$.
Consequently, careful analysis of line profiles in both the wavelength and Fourier 
space, combined by extensive theoretical simulations, may open the way to clarifying the 
rotational properties of sharp-line stars, though this would be a much more difficult 
task compared to the case of broad-line stars. 
 
\section{Summary and conclusion}

The nature of differential rotation in A-, F-, and G-type stars has been 
extensively investigated by Reiners et al. by making use of the fact 
that the ratio of the first- and second-zero frequencies ($q_{2}/q_{1}$) in the 
Fourier transform of spectral line profiles is a useful indicator. 

However, their approach relies on the conventional formulation that 
the observed flux profile is expressed as a convolution of the rotational 
broadening function and the intrinsic profile (cf. equation(1)), which implicitly 
assumes that the local intensity profile does not change over the disk. 
Although this postulation of profile constancy is unrealistic in the strict sense, 
how the result is affected by this approximation has barely been investigated so far.

In order to examine the validity of this assumption, flux profiles were simulated 
for eight fictitious lines (showing different center-to-limb variations) for various 
combinations of $v_{\rm e}\sin i$, $\alpha$, and $i$, while integrating the intensity 
profiles at many points on the visible disk calculated by using a model atmosphere 
corresponding to a mid-F dwarf along with adequately specified angle-dependent 
$v_{\rm mac}(\theta)$ and $v_{\rm mic}(\theta)$.
Fourier transforms of these profiles were further computed in order to check 
whether the zeros are detected at the positions predicted by the rotational 
broadening function.
 
For this purpose, a large grid of standard rotational broadening functions (for 
various combinations of $\alpha$, $i$, and $\epsilon$) were calculated in advance 
by using the iso-velocity contours, and the zero frequencies were measured from 
their Fourier transforms. These data are presented as online materials.

The check revealed that the situation critically depends on the extent of $v_{\rm e}\sin i$.
In case of $v_{\rm e}\sin i \gtsim 20$~km~s$^{-1}$ where rotational broadening 
is predominant over other broadening velocities (several km~s$^{-1}$),
the first/second zeros of the transform are detected almost at the expected positions.
In contrast, deviations begin to appear as $v_{\rm e}\sin i$ is lowered,  
and the zero features of the transform are totally different from
the expectation at $v_{\rm e}\sin i$ as low as $\sim 10$~km~s$^{-1}$, 
which means that the classical formulation is not valid any more.   
In summary, while zero-frequency approach is safely applicable to studying 
differential rotation in the former broader-line case, it would be difficult 
to practice for the latter sharp-line case.

\appendix

\section{Anisotropy of macro- and micro-turbulences in the solar photosphere}

The heart of this investigation is to simulate the profiles of various lines
at different disk points of an F5 dwarf as realistic as possible.  
For this purpose, the solar center--limb variation data in the strengths as well as 
widths of many spectral lines recent published by Takeda and UeNo (2019) were
consulted for reference, since we may reasonably postulate that the qualitative 
characteristics of the surface properties are not much different for F5 and G2 dwarfs.

As usually done in stellar spectroscopy, a very rough modeling was exploited 
for the solar photospheric velocity fields (affecting line profiles), which are 
divided into``micro''-turbulence ($v_{\rm mic}$) and ``macro''-turbulence ($v_{\rm mac}$) 
and separately treated, where the former (microscopic scale) is included in the Doppler 
with of the line-opacity profile (like thermal velocity) while the latter (macroscopic 
scale) acts as a global velocity distribution function (like rotational broadening 
function) to be convolved with the intrinsic profile. Although this dichotomous model
characterized by two parameters ($v_{\rm mic}$ and $v_{\rm mac}$) is known to be far 
from realistic (especially compared with the recent state-of-the-art 3D time-dependent 
hydrodynamical modeling), it is very useful in the practical sense because the widths
and strengths of any spectral lines can be reasonably modeled if these fudge 
parameters are appropriately adjusted. 

It has been known that angle-dependence has to be introduced to both $v_{\rm mic}$ and 
$v_{\rm mac}$ in order to reproduce the observed solar center-limb variations of spectral 
line strengths/widths, as occasionally reported by previous investigators; e,g., 
Holweger, Gehlsen, and Ruland (1978) for the microturbulence, or Gurtovenko (1976; cf. 
Fig.~2 therein) for the macroturbulence (non-thermal velocity dispersion). 
Actually, the same argument could be made based on the data compiled by Takeda 
and UeNo (2019), which eventually lead to the use of equation~(10) (for $v^{\odot}_{\rm mac}$) 
and equation~(12) (for $v^{\odot}_{\rm mac}$) adopted in subsection~3.3.  
Since the use of such anisotropic $v_{\rm mic}$ or $v_{\rm mac}$ is essentially important 
for the conclusion of this study (cf. appendix~2), some additional explanations are presented 
regarding the validity of these relations. 

Figure~9 depicts how the $v_{\rm mac}$\footnote{Note that this parameter was denoted as 
$V_{\rm los}$ in Takeda and UeNo (2019).} values depend upon the mean formation depth 
($\langle \log \tau \rangle$) and the view angle ($\sin\theta$), which were derived by 
Takeda and UeNo (2019)  at each point of the solar disk for 280 Fe~{\sc i} lines, 
Figures 9e--h manifestly show that $v_{\rm mac}$ progressively increases toward the limb
and its $\theta$-dependence can be well represented by equation~(10) on the average.
Since it appears difficult to attribute this tendency solely to the depth-dependence 
as was occasionally argued in the old studies, some kind of real anisotropy should 
exist in the macroscopic turbulent velocity dispersion. 

Regarding $v_{\rm mic}$, it is known to significantly affect the strengths
of stronger saturated lines (on the flat part of the curve of growth) but not 
those of weak lines (on the linear part of the curve of growth).
In figures~10a--e are plotted the abundance differences relative to the disk-center value 
against $\cos \theta$, which were derived for 280 Fe~{\sc i} lines based on $W$ values 
published by Takeda and UeNo (2019) by assuming $v_{\rm mic} = 1$~km~s$^{-1}$. 
It can be seen from these figures that (although the consistency is almost accomplished 
for weak lines) the abundance discrepancy for stronger lines progressively increases 
towards the limb, which indicates that the observed center--limb trends of stronger lines
can not be reproduced by a constant microturbulence and thus an angle-dependent $v_{\rm mic}$ 
increasing toward the limb is necessary to achieve consistency (i.e., position-independent 
abundances). Actually, figure~10f reveals that the systematic discrepancy can be considerably 
mitigated by using the $\theta$-dependent $v_{\rm mic}$ given by equation~(12).  

\section{Importance of the angle-dependence of turbulent velocities}

With an aim to simulate the profiles of various lines at different disk points 
of an F5 dwarf as realistic as possible, this study adopted special $\theta$-dependent 
anisotropic forms for both $v_{\rm mic}$ and $v_{\rm mac}$ expressed by equations 
(11) and (13) as described in subsection~3.3, which were devised in analogy with 
the solar case by making use of Takeda and UeNo's (2019) recent study (cf. appendix~1).
However, it may be more common for most people to simply assign constant values to 
these fudge parameters for modeling line profiles if information is lacking. 
In this respect, it would be instructive to examine how the conclusions derived in 
subsection~4.2 are affected if constant (angle-independent)
values are used for $v_{\rm mic}$ or $v_{\rm mac}$. 

For this purpose, additional test calculations were carried out for the following 
three cases with different treatments of $v_{\rm mic}$ and $v_{\rm mac}$:
Case 1 ... $v_{\rm mic} = 1.6$~km~s$^{-1}$ (constant) while $v_{\rm mac}(\theta)$ unchanged.
Case 2 ... $v_{\rm mic}(\theta)$ unchanged while $v_{\rm mac} = 2.4$~km~s$^{-1}$ (constant).
Case 3 ... $v_{\rm mic} = 1.6$~km~s$^{-1}$ (constant) and $v_{\rm mac} = 2.4$~km~s$^{-1}$ (constant).
The resulting behaviors of $q_{2}/q_{1}$ corresponding to $v_{\rm e}\sin i = 10$~km~s$^{-1}$ 
are illustrated in figure~11 (left, middle, and right panels correspond to Cases 1, 2, and 3,
respectively), which are so arranged as to be directly comparable with the left-hand panels 
in figure~7 ($v_{\rm e}\sin i = 10$~km~s$^{-1}$ case, where the measured ratios are in marked 
conflict with the expected value from the broadening function). The following characteristics 
are read by inspecting figure~11 in comparison with figure~7.
\begin{itemize}
\item
Case 1 results (left panels in figure~11) indicate that changing only $v_{\rm mic}$ 
(while $v_{\rm mac}$ still $\theta$-dependent) does not 
have any essential impact, While this is naturally expected for weak lines 
($W_{\mu=1} = 30$~m\AA; lower 4 panels) which are anyhow insensitive to $v_{\rm mic}$), 
the situation is not much improved even for $v_{\rm mic}$-sensitive stronger lines 
($W_{\mu=1} = 100$~m\AA; upper 4 panels) either.
\item 
However, the results of Case 2 (middle panels in figure~11) reveal that adopting a constant 
$v_{\rm mac}$ (while $v_{\rm mic}$ still $\theta$-dependent) yields a remarkable 
improvement (i.e., recovery of expected $q_{2}/q_{1}$ ratios) for three weak lines 
(2600c00w030, 2600c30w030, 2600c60w030), though the other remaining lines still show 
appreciable discrepancy.
\item
Interestingly, expected $q_{2}/q_{1}$ ratios are much better (though not perfectly) reproduced 
for all of the 8 lines in Case 3 (right panels in figure~11) where constant values are used for 
both $v_{\rm mic}$ (1.6~km~s$^{-1}$) and $v_{\rm mac}$ (2.4~km~s$^{-1}$), which means that 
application of equation~(1) is not necessarily bad even at $v_{\rm e}\sin i$ as low as 
$\sim 10$~km~s$^{-1}$ (i.e., in marked contrast to the conclusion of this study).
\item 
Consequently, realistic treatment (i.e., inclusion of $\theta$-dependence) for $v_{\rm mic}$ 
as well as $v_{\rm mac}$ is essentially important, because simply adopting constant values
for these parameters yields incorrect results. This argument applies more significantly to 
$v_{\rm mac}$ (controlling widths of all lines) than to $v_{\rm mic}$ (affecting the 
strengths of stronger lines). 
\end{itemize}

\setcounter{table}{0}
\begin{table}[h]
\small
\caption{Fictitious test lines used for the simulation.}
\begin{center}
\begin{tabular}{ccccccc}\hline\hline
Code & $\lambda$ & Species & $\chi_{\rm low}$ & $\log\epsilon gf$ & $W_{0}$ & $W_{\rm limb}$ \\
(1)  &  (2) & (3) & (4) & (5) & (6) & (7)\\
\hline
2600c00w100 &  5000 & Fe~{\sc i} & 0.0 & 3.98 & 100 & 118 \\
2600c30w100 &  5000 & Fe~{\sc i} & 3.0 & 6.64 & 100 & 102 \\
2600c60w100 &  5000 & Fe~{\sc i} & 6.0 & 8.91 & 100 &  74 \\
0800c90w100 &  5000 & O~{\sc i}  & 9.0 & 8.78 & 100 &  51 \\
2600c00w030 &  5000 & Fe~{\sc i} & 0.0 & 2.64 &  30 &  48 \\
2600c30w030 &  5000 & Fe~{\sc i} & 3.0 & 5.25 &  30 &  37 \\
2600c60w030 &  5000 & Fe~{\sc i} & 6.0 & 7.72 &  30 &  25 \\
0800c90w030 &  5000 & O~{\sc i}  & 9.0 & 7.58 &  30 &  11 \\
\hline
\end{tabular}
\end{center}
(1) Line code. (2) Wavelength (in \AA). (3) Element species. (4) Lower excitation potential (in eV).
(5) Logarithm of the product of adopted oscillator strength ($gf$) and elemental abundance ($\epsilon$)
which was so adjusted as to reproduce the specified equivalent width at the disk center,
where $\log\epsilon$ is defined as $\log (N_{\rm X}/N_{\rm H}) + 12$ as usual. (6) Equivalent width 
(in m\AA) at the disk center ($\mu = \cos\theta = 1.0$).  (7) Equivalent width (in m\AA)
at the limb ($\mu = \cos\theta = 0.1$).  
\end{table}

\onecolumn

\newpage

\setcounter{figure}{0}
\begin{figure}
\begin{center}
  \FigureFile(80mm,80mm){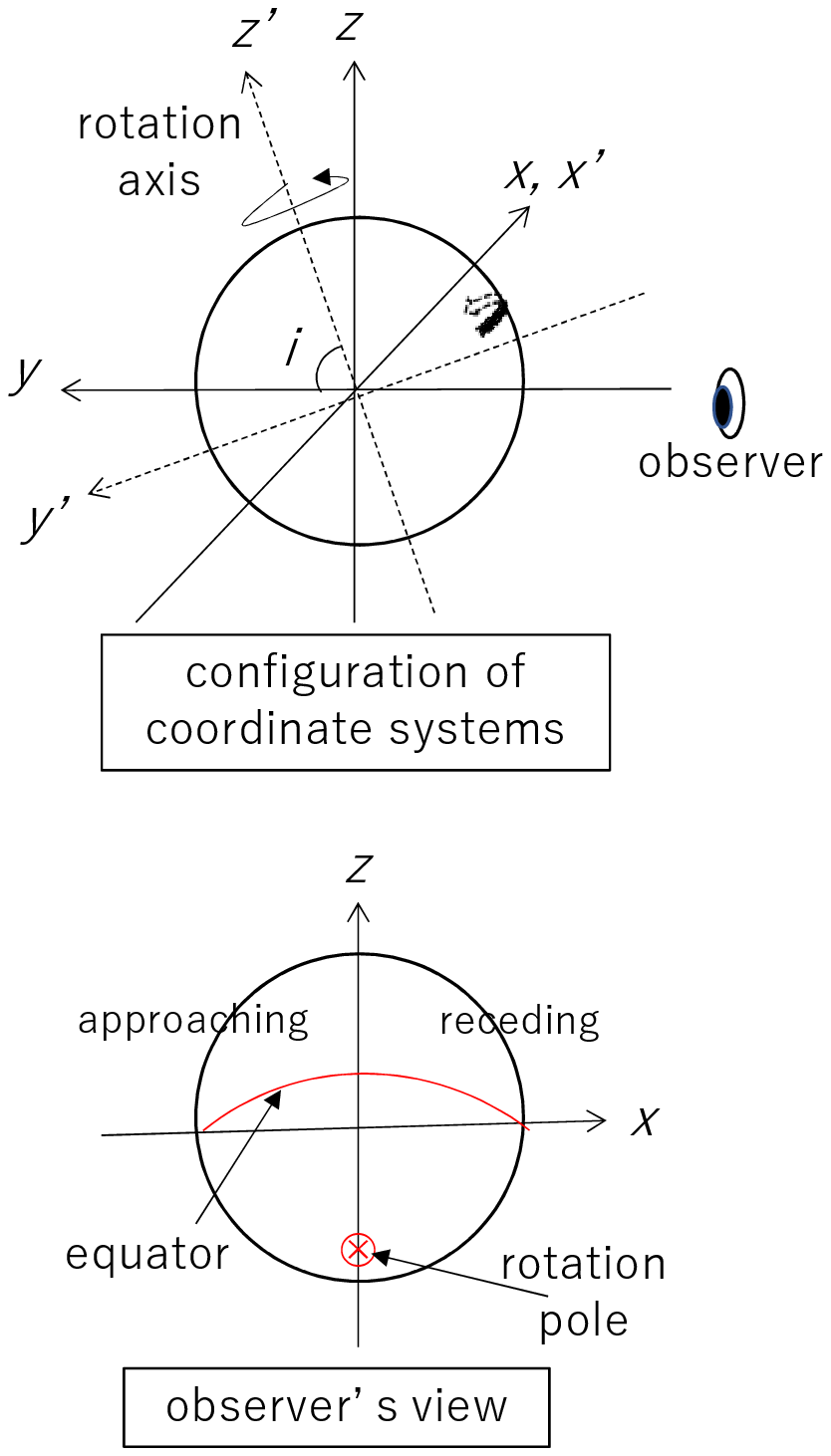}
\end{center}
\caption{
Upper figure: Schematic description of two coordinate systems ($x,y,z$) and ($x',y',z'$), 
where the latter is rotated by an angle of $90^{\circ} -i $ ($i$ is the inclination angle of rotation) 
relative to the former (around the common $x$- and $x'$-axis) so that the $y$-axis matches 
the observer's line of sight and the $z'$-axis coincides with the rotational axis. 
Lower figure: Observer's view of the stellar disk, which is the projection
of the stellar sphere onto the $x$--$z$ plane.  
}
\label{fig:1}
\end{figure}

\setcounter{figure}{1}
\begin{figure}
  \begin{center}
    \FigureFile(130mm,170mm){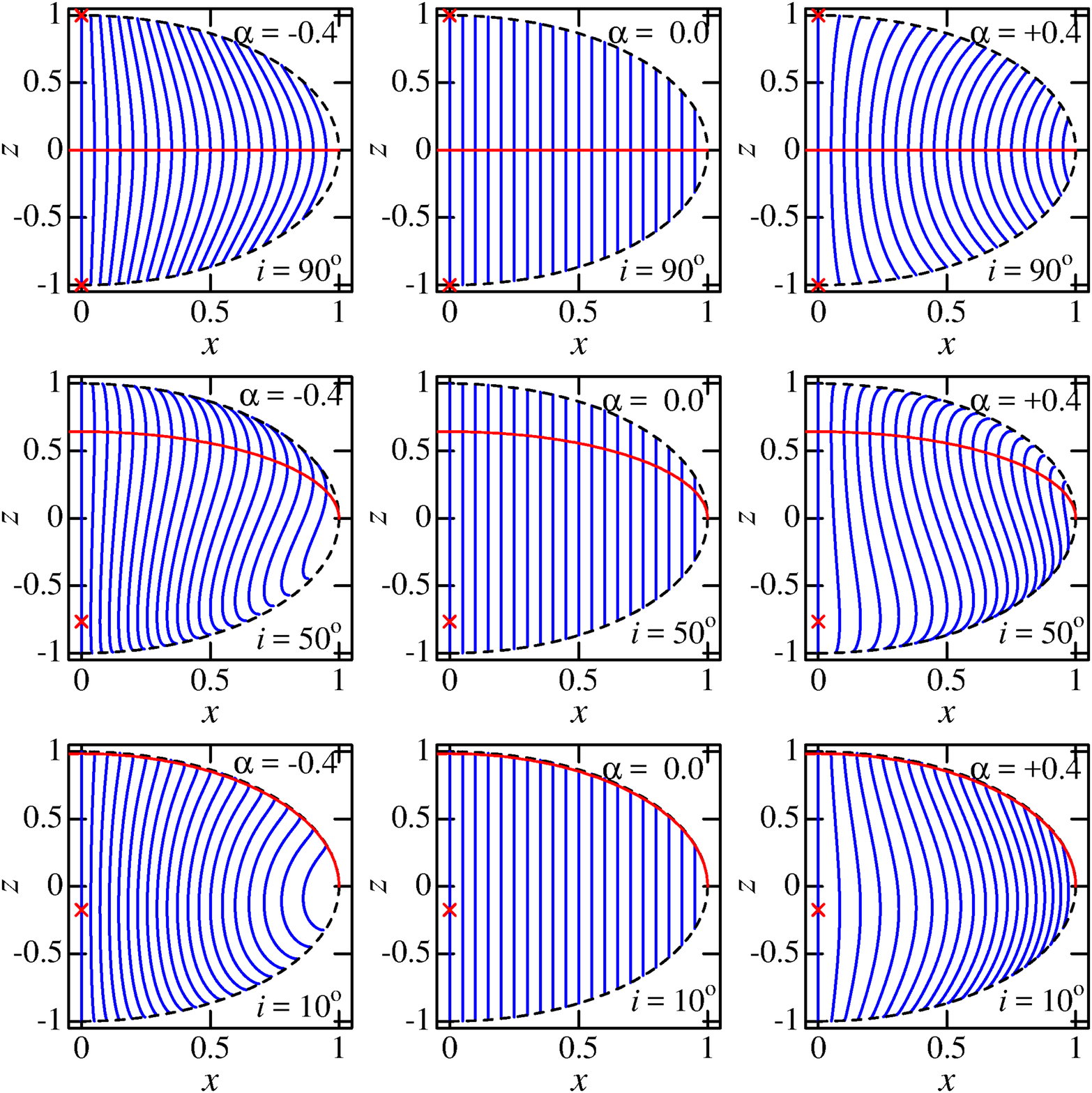}
  \end{center}
\caption{
Contours of iso-velocity (i.e., constant radial velocity) on the stellar disk corresponding to
$k = 0.00, 0.05, 0.10, \cdots 0.90, 0.95$, where $k (\equiv v/v_{\rm e}\sin i)$
is the non-dimensional parameter, computed for representative combinations
of $\alpha$ and $i$. The left, center, and right panels are for $\alpha = -0.4$,
$0.0$, and $+0.4$, while the top, middle, and bottom panels are for $i = 90^{\circ}$,
$50^{\circ}$, and $10^{\circ}$, respectively. Also indicated are the equator (red line) 
and the rotation pole (red cross) visible on the disk. 
}
\end{figure}

\setcounter{figure}{2}
\begin{figure}
  \begin{center}
    \FigureFile(130mm,170mm){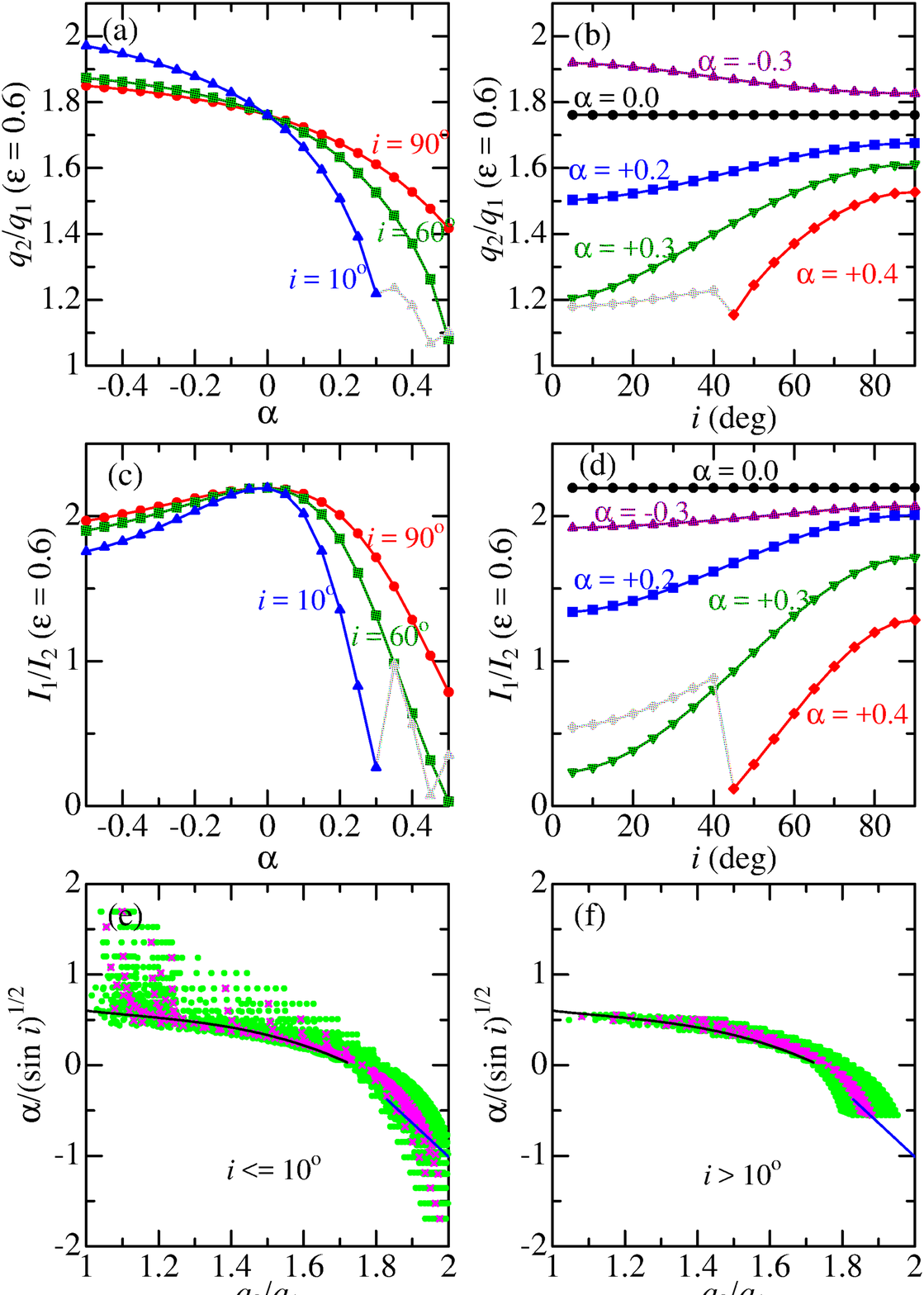}
  \end{center}
\caption{
In the upper panels are plotted the $q_{2}/q_{1}$ values (second-to-first zero 
frequency ratio) against (a) $\alpha$ and (b) $i$, while the middle
panels (c) and (d) similarly show the runs of $I_{1}/I_{2}$ (first-to-second 
sidelobe height ratio), which correspond to the limb-darkening coefficient of $\epsilon = 0.6$ 
and are to be compared with Reiners and Schmitt's (2002) Fig.~5 and Fig.~7.
Note that the discontinuous phenomena seen for the case of high $\alpha$ and low $i$
(obscured in gray) are due to the fact that the first zero in the expected  
position is no more detected (and thus the ``to-be'' second zero is in turn 
regarded as the first zero).  The $\alpha/\sqrt{\sin i}$ vs. $q_{2}/q_{1}$ relations
are illustrated in the lower panels (e) (for $i \le 10^{\circ}$) and (f) (for 
$i > 10^{\circ}$), where all data and those only for $\epsilon = 0.6$ are depicted 
by filled circles and crosses, respectively, which should be compared with Fig.~11 
of Reiners and Schmitt (2003) [the solid line shows the analytical relation represented
by equations~(5) and (6) of their paper]. 
}
\end{figure}

\setcounter{figure}{3}
\begin{figure}
  \begin{center}
    \FigureFile(80mm,100mm){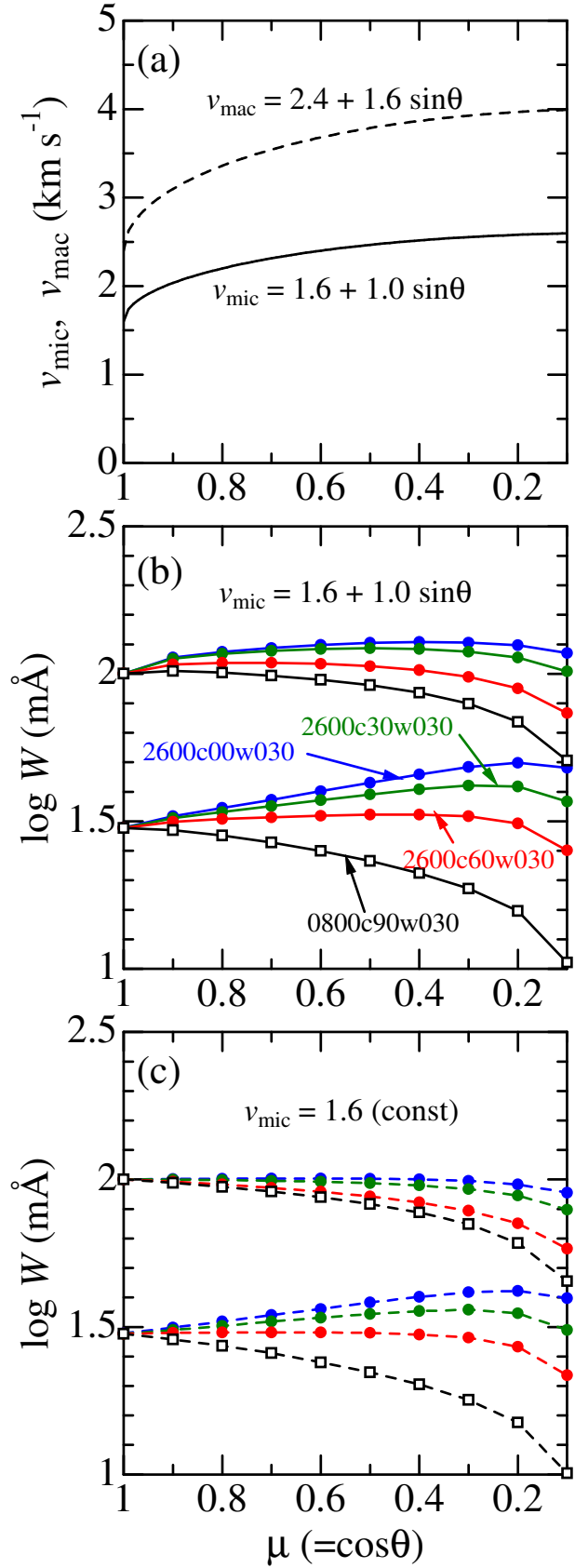}
  \end{center}
\caption{
(a) Adopted $v_{\rm mic}(\theta)$ (microturbulence given by equation~(13); solid line) and 
$v_{\rm mac}(\theta)$ (macroturbulence given by equation~(11); dashed line) plotted against 
$\mu (\equiv \cos\theta)$.
(b) Angle-dependence of the equivalent widths ($W$) computed by using $v_{\rm mic}(\theta)$
of equation~(13) for 8 fictitious lines. (Note that the line data listed in table~1 are 
arranged in the same order as the curves in this figure.)
(c) Angle-dependence of the equivalent widths ($W$) for 8 fictitious lines similar to (b),
but computed by using the constant $v_{\rm mic}$ of 1.6~km~s$^{-1}$ 
}
\end{figure}

\setcounter{figure}{4}
\begin{figure}
  \begin{center}
    \FigureFile(80mm,100mm){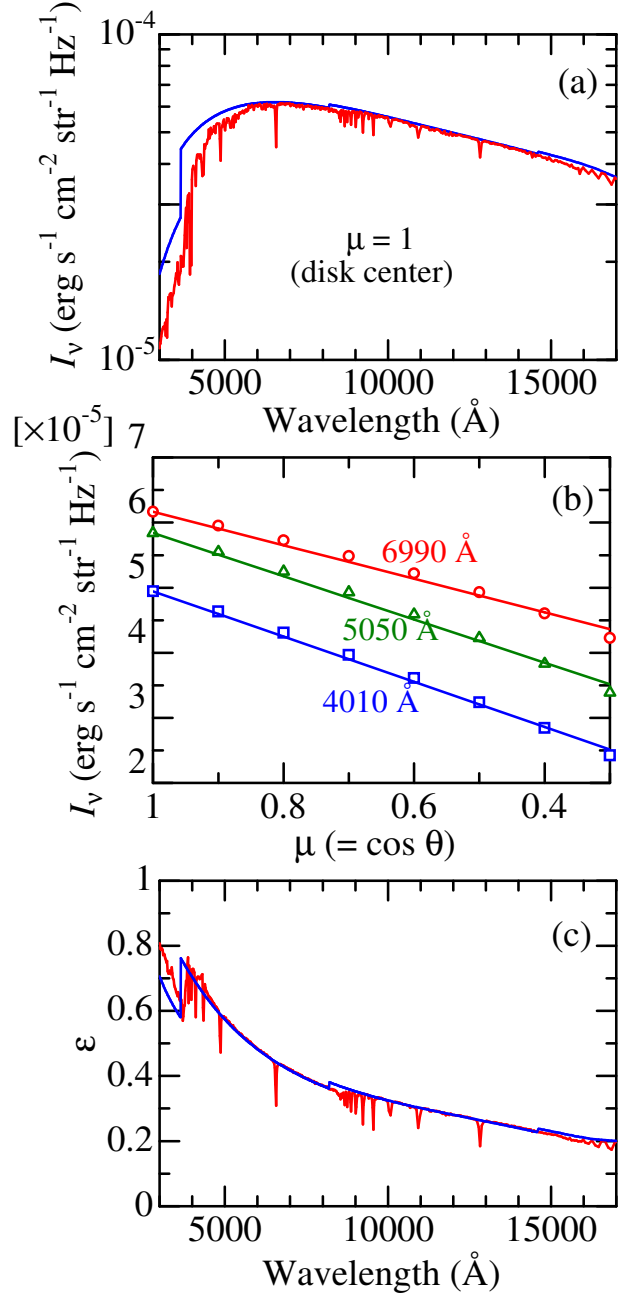}
  \end{center}
\caption{
(a) Energy distribution of the specific intensity ($I_{\nu}$) at the line center 
($\mu = \cos\theta = 1$)
computed from the atmospheric model ($T_{\rm eff} = 6500$~K, $\log g = 4.29$,
solar abundance) adopted in this study, where the line-included and pure-continuum
distributions are depicted in red and blue lines, respectively. 
(b) Angle-dependence of continuum $I(\nu)$ at 4010~\AA, 5050~\AA, and 6990~\AA\
(symbols) along with the linear-regression lines derived from these data between
$\mu = 1$ and 0.3 (lines).
(c) Wavelength-dependence of $\epsilon$ (limb-darkening coefficient)
defined as the slope of the linear-regression line such like as in panel (b). 
As in panel (a), the red and blue lines correspond to the cases of line-included 
intensity and pure-continuum intensity, respectively.
}
\end{figure}

\setcounter{figure}{5}
\begin{figure}
  \begin{center}
    \FigureFile(120mm,170mm){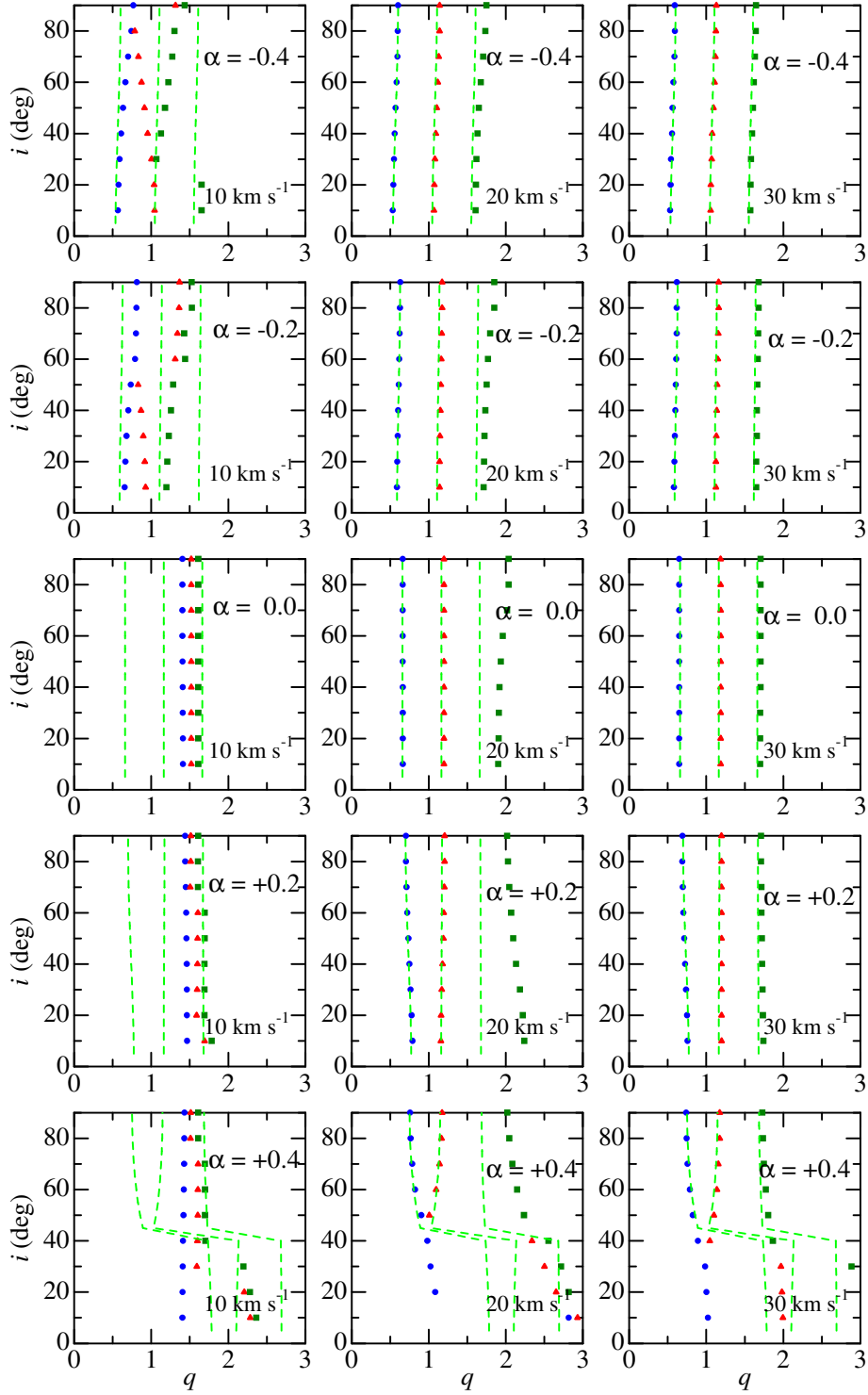}
  \end{center}
\caption{
Graphical display showing how the first-, second-, and third-zero frequencies
(depicted by circles, triangles, and squares, respectively) 
computed for the 2600c30w100 line behave with a change in $i$ 
(inclination angle) for different cases of $\alpha$ and $v_{\rm e}\sin i$.
Note that the non-dimensional frequency $q$ [cf. equation~(14)] is used in 
the abscissa instead of the actual $\sigma$ (\AA$^{-1}$), in order to enable a direct comparison 
irrespective of $v_{\rm e}\sin i$. The zero positions expected for the standard rotational 
broadening function ($\epsilon = 0.6$; cf. subsection~4.1) are shown by dashed lines. 
The panels in the first, second, third, fourth, and fifth row correspond to $\alpha = -0.4$,
$-0.2$, $0.0$, $+0.2$, and $+0.4$, while those in the left, center, and right column
are for $v_{\rm e}\sin i = 10$, 20, and 30~km~s$^{-1}$, respectively. 
}
\end{figure}

\setcounter{figure}{6}
\begin{figure}
  \begin{center}
    \FigureFile(130mm,170mm){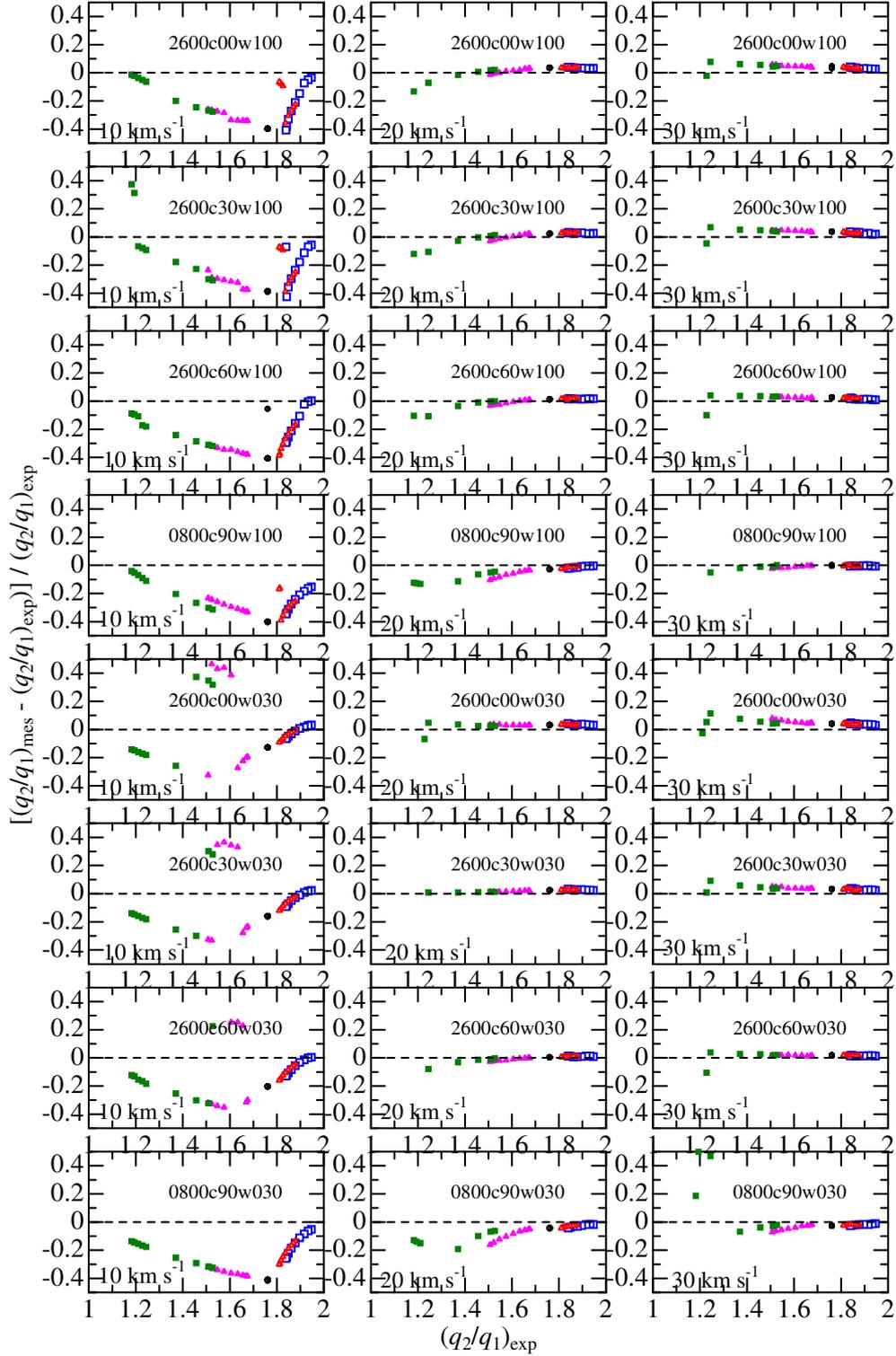}
  \end{center}
\caption{
Diagrams showing how the measured second-to-first zero frequency ratios 
[$(q_{2}/q_{1})_{\rm mes}$] deviate from the expected values from 
the standard rotational broadening function [$(q_{2}/q_{1})_{\rm exp}$; 
corresponding to $\epsilon = 0.6$], computed for the eight lines presented 
in table~1 (note that the panels in each of 
the eight rows are arranged in the same order as in the table).
The left, middle, and right panels are for $v_{\rm e}\sin i = 10$, 20, and 
30~km~s$^{-1}$, respectively. The results (for nine $i$ values from
10$^{\circ}$ to 90$^{\circ}$) for each of the five different $\alpha$ values 
are shown here, which are discriminated by the symbols:  $\alpha = -0.4$
(open squares), $-0.2$ (open triangles), $0.0$ (half-filled circles),
$+0.2$ (filled triangles), and $+0.4$ (filled squares).  
}
\end{figure}

\setcounter{figure}{7}
\begin{figure}
  \begin{center}
    \FigureFile(130mm,160mm){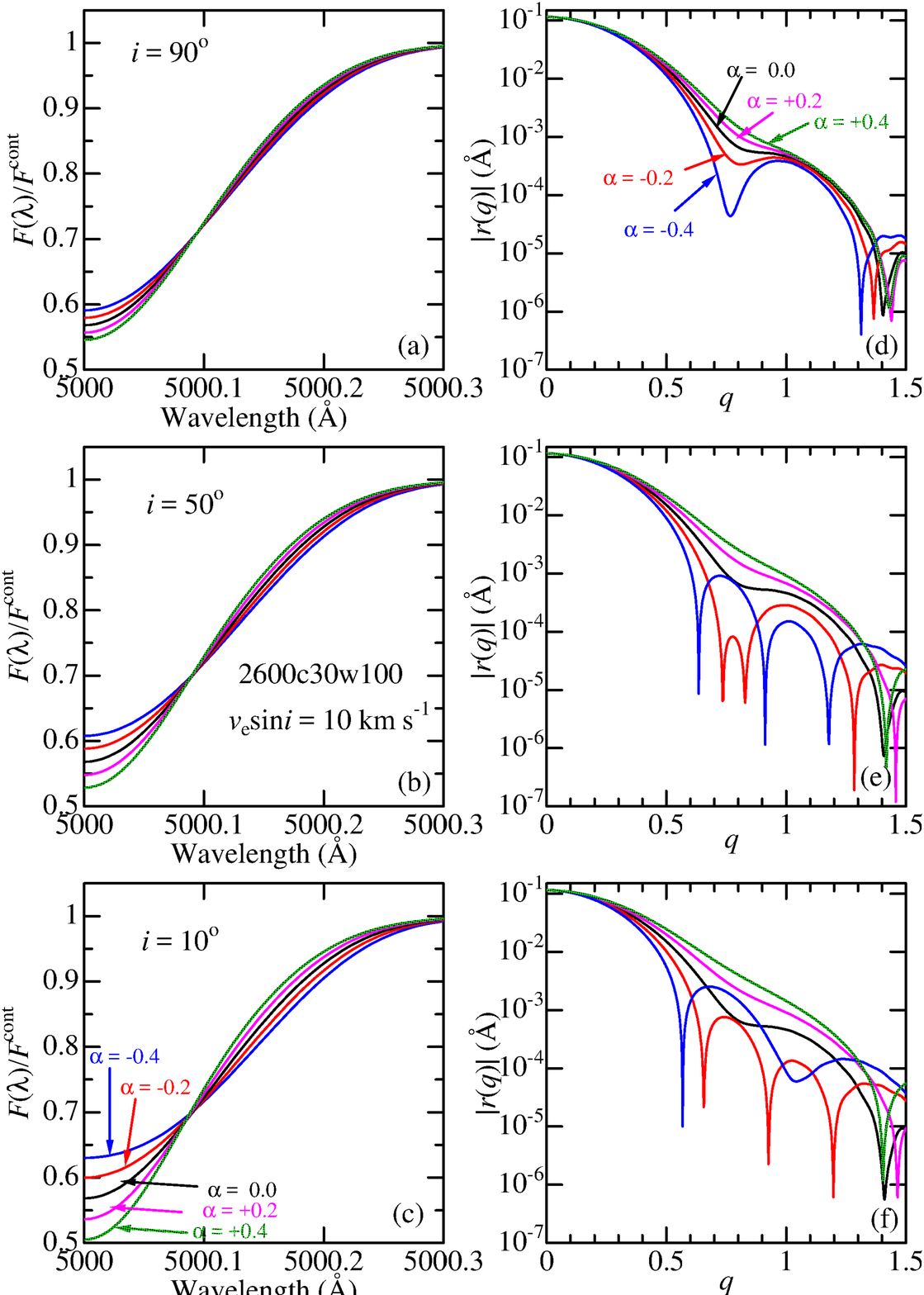}
  \end{center}
\caption{
The left panels display the residual fluxes [$F(\lambda)/F^{\rm cont}$] of the 
2600c30w100 line simulated with the specified $v_{\rm e}\sin i$ of 10~km~s$^{-1}$
for the five values of $\alpha$ ($-0.4$, $-0.2$, 0.0, +0.2, and $+0.4$),
while the right panels  show the corresponding Fourier transform amplitudes [$|r(q)|$]  
of the line depth [$R(\lambda) \equiv 1 - F(\lambda)/F^{\rm cont}$] (where 
non-dimensional $q$ is used as abscissa such as like in figure~6): (a,d) $i=90^{\circ}$, 
(b,e) $i=50^{\circ}$, and (c, f) $i=10^{\circ}$.
}
\end{figure}

\setcounter{figure}{8}
\begin{figure} 
  \begin{center}
    \FigureFile(130mm,160mm){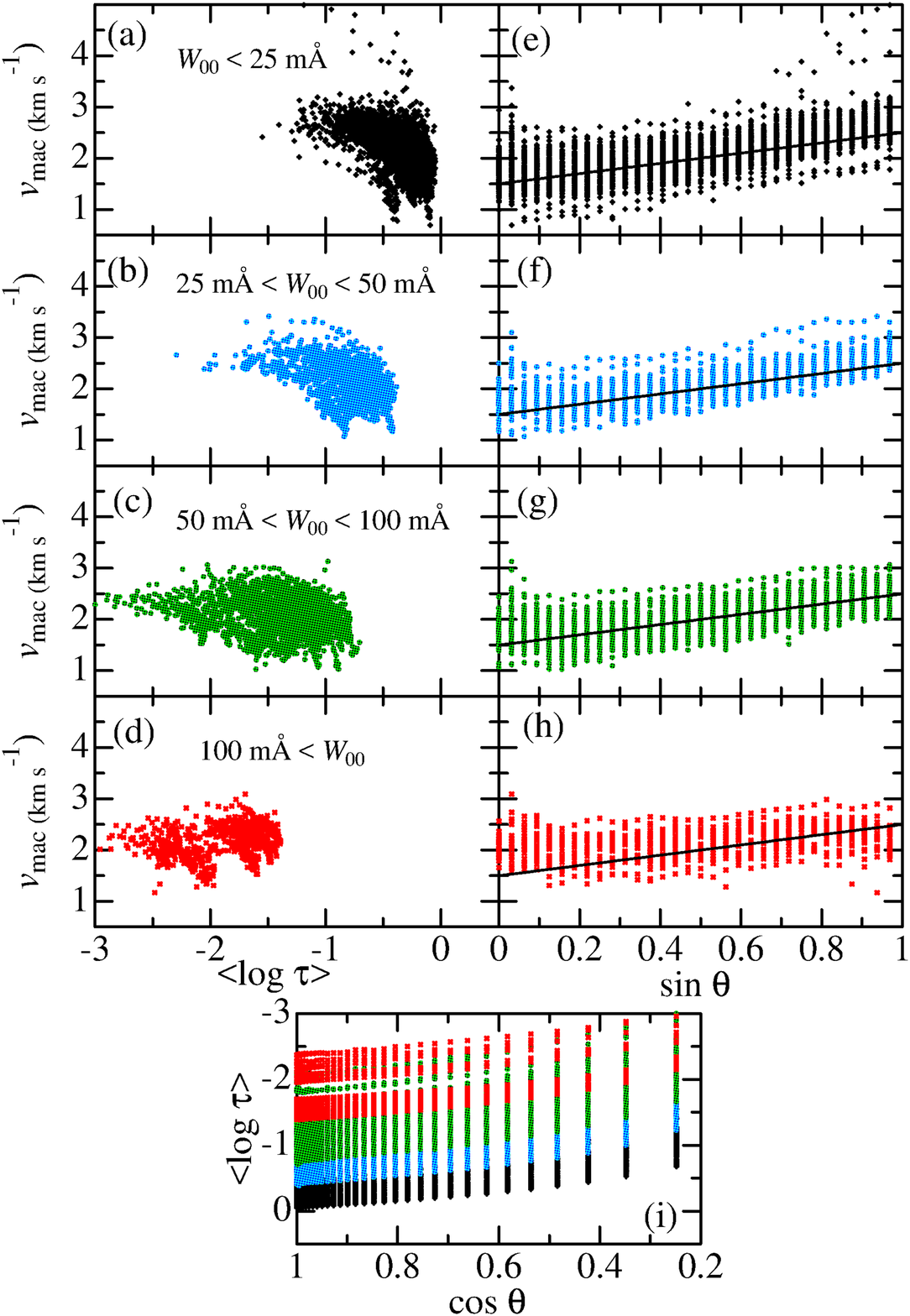}
  \end{center}
\caption{
The $v_{\rm mac}$ values (line-of-sight velocity dispersion of Gaussian 
macroturbulence) at each point of the solar disk, which were derived by 
Takeda and UeNo (2019) for 280 Fe~{\sc i} lines, are plotted against 
$\langle \log\tau \rangle$ (mean formation depth; left panels (a)--(d)) 
and $\sin\theta$ (right panels (e)--(h)), where the data in each panel 
are grouped according to the range of disk-center equivalent width ($W_{00}$).
(a),(e): $W_{00} <$~25~m\AA\ (black symbols), 
(b),(f): 25~m\AA~$\le W_{00} <$~50~m\AA\ (blue symbols),
(c),(g): 50~m\AA~$\le W_{00} <$~100~m\AA\ (green symbols), and
(d),(h): 100~m\AA~$\le W_{00}$ (red symbols).
The adopted mean relation, $v_{\rm mac} = 1.5 + 1.0\sin \theta$ (cf. equation~(10)),
is drawn by solid lines in each of the right panels (e)--(h). 
The bottom panel (i) shows the correlation between $\langle \log\tau \rangle$ 
and $\cos\theta$ (all data overplotted).
}
\end{figure}

\setcounter{figure}{9}
\begin{figure} 
  \begin{center}
    \FigureFile(130mm,160mm){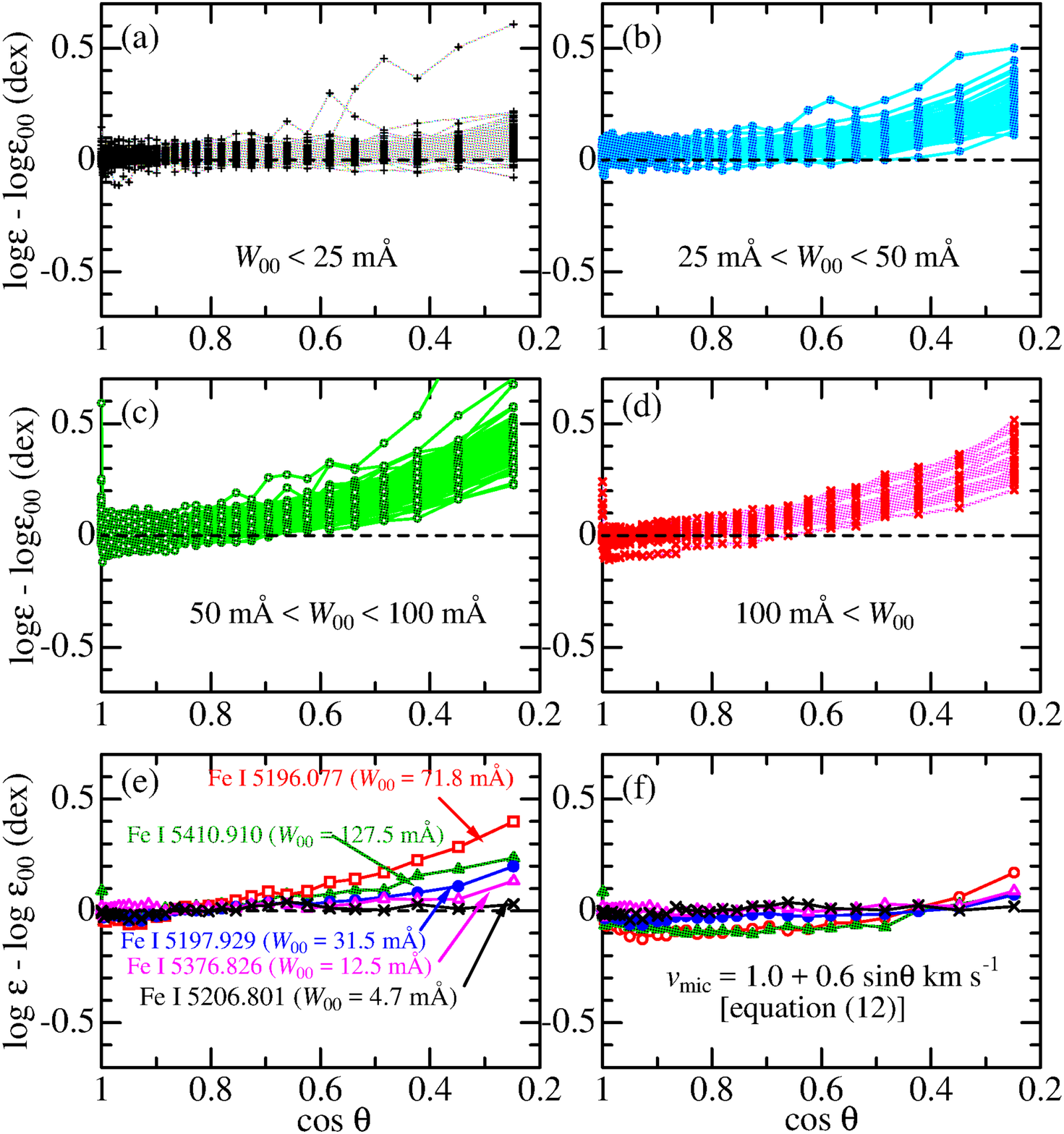}
  \end{center}
\caption{
Differences of abundances relative to the disk center-value
($\log\epsilon - \log\epsilon_{00}$), which were derived based on the 
solar center--limb equivalent widths ($W$) of 280 Fe~{\sc i} lines 
published by Takeda and UeNo (2019) by assuming a constant microturbulence
of $v_{\rm mic} = 1$~km~s$^{-1}$ (except for panel (f)), are plotted against 
$\cos\theta$,
where each of panels (a), (b), (c), and (d) correspond to different four 
line-strength classes as in figure~9 (with the same colors).
The results for five representative lines of different strengths
are shown in the bottom panels (e) and (f), each corresponding to  
$v_{\rm mic} = 1$~km~s$^{-1}$ and angle dependent $v_{\rm mic}$ 
($1.0 + 0.6\sin \theta$~km~s$^{-1}$ given by equation~(12)), respectively: 
Fe~{\sc i} 5206.801 ($W_{00} =   4.7$~m\AA),
Fe~{\sc i} 5376.826 ($W_{00} =  12.5$~m\AA),
Fe~{\sc i} 5197.929 ($W_{00} =  31.5$~m\AA),
Fe~{\sc i} 5196.077 ($W_{00} =  71.8$~m\AA), and 
Fe~{\sc i} 5410.910 ($W_{00} = 127.5$~m\AA).
}
\end{figure}

\setcounter{figure}{10}
\begin{figure}
  \begin{center}
    \FigureFile(130mm,170mm){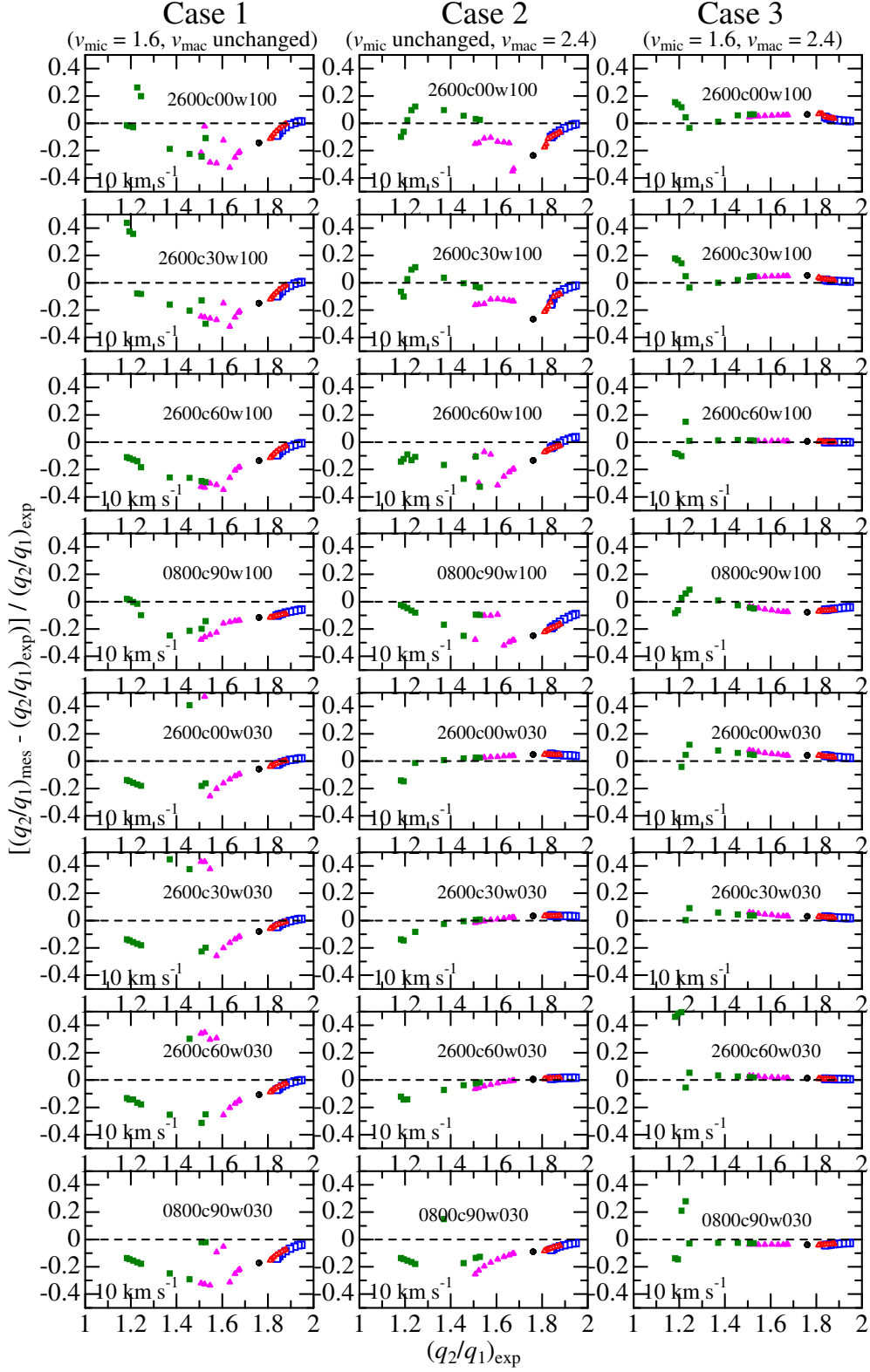}
  \end{center}
\caption{
Results of test calculations showing how the serious deviations of measured 
$(q_{2}/q_{1})_{\rm mes}$ from the expected $(q_{2}/q_{1})_{\rm mes}$ for 
the $v_{\rm e}\sin i = 10$~km~s$^{-1}$ cases are changed by using the simple 
constant $v_{\rm mac}$ (2.4~km~s$^{-1}$) or $v_{\rm mic}$ (1.6~km~s$^{-1}$) 
instead of the standard $\theta$-dependent relations [cf. equations (11) and (13)]. 
Left panels --- Case 1 ($v_{\rm mac}(\theta)$ unchanged  
while $v_{\rm mic}$ = 1.6~km~s$^{-1}$).
Middle panels --- Case 2 ($v_{\rm mac}$ = 2.4~km~s$^{-1}$ while $v_{\rm mic}
(\theta)$ unchanged).
Right panels --- Case 3 ($v_{\rm mac}$ = 2.4~km~s$^{-1}$ and 
$v_{\rm mic}$ = 1.6~km~s$^{-1}$).
These figures panels are so arranged as to be directly comparable with the 
left-hand panels of figure~7 ($v_{\rm e}\sin i = 10$~km~s$^{-1}$ case, 
corresponding to the standard $v_{\rm mic}(\theta)$ and $v_{\rm mic}(\theta)$);
see the caption therein for more details. 
}
\end{figure}

\end{document}